\documentclass[twocolumn,aps,prl,noshowpacs,amsmath,amssymb,superscriptaddress,letterpaper]{revtex4}
\pdfoutput=1

\usepackage{amsmath}
\usepackage{amsfonts}
\usepackage{graphicx}
\usepackage{hyperref}

\DeclareGraphicsExtensions{.pdf}

\newcommand{\Hilbert}{\mathcal{H}}
\newcommand{\dHilbert}{d}
\newcommand{\LieA}{\mathcal{L}}
\newcommand{\LieAC}{\LieA_{C}}
\newcommand{\HDesire}{H_{d}}
\newcommand{\UDesire}{U_{d}}
\newcommand{\HInternal}{H_{int}}
\newcommand{\UInternal}{U_{int}}
\newcommand{\HCalc}{H_{calc}}
\newcommand{\PX}[1]{\sigma_{\text{x}}^{(#1)}}
\newcommand{\PY}[1]{\sigma_{\text{y}}^{(#1)}}
\newcommand{\PZ}[1]{\sigma_{\text{z}}^{(#1)}}
\newcommand{\RP}{\rho_{+}}
\newcommand{\RM}{\rho_{-}}
\newcommand{\RI}{\rho_{i}}
\newcommand{\RF}{\rho_{f}}
\newcommand{\UP}{U_{+}}
\newcommand{\UM}{U_{-}}
\newcommand{\GU}{\mathcal{G}_{\mathcal{U}}}
\newcommand{\GH}{\mathcal{G}_{\mathcal{H}}}
\newcommand{\DU}{\mathcal{D}_{U}}	
\newcommand{\diag}[1]{\mathrm{diag}(#1)}
\newcommand{\diagb}[3]{\mathrm{diag}(#1)_{#2}^{#3}}
\newcommand{\tnot}{t_{0}}

\newcommand{\up}{\uparrow}
\newcommand{\down}{\downarrow}
\newcommand{\pa}{\up \down \down \down}
\newcommand{\pb}{\down \down \down \down}
\newcommand{\pc}{\down \down \down \up}

\newcommand{\conj}[2]{{#2}{#1}{#2^{\dagger}}}
\newcommand{\jnoc}[2]{{#2^{\dagger}}{#1}{#2}}

\newcommand{\GenLin}[1]{\mathrm{GL}({#1})}
\newcommand{\genLin}[1]{\mathfrak{gl}({#1})}
\newcommand{\skewHerm}[1]{\mathfrak{u}({#1})}
\newcommand{\Unitary}[1]{\mathcal{U}(#1)}
\newcommand{\linearOp}[1]{\mathcal{B}(#1)}
\newcommand{\GenLinBlock}[2]{\mathrm{GL}({#1},\dots,{#2})}

\newcommand{\UnitaryBlock}[2]{\mathcal{U}(#1,\dots,{#2})}
\newcommand{\centralizer}[1]{\mathcal{C}_{GL}(#1)}
\newcommand{\project}[2]{\mathcal{P}_{#2}(#1)}

\newcommand{\ket}[1]{\left| #1 \right>}
\newcommand{\bra}[1]{\left< #1 \right|}

\DeclareMathOperator*{\argmin}{argmin}
\DeclareMathOperator*{\argmax}{argmax}

\begin{document}

\title{Equivalent Hamiltonians for State to State Transfer \\in the Case of Partial Quantum Control}
\author{I. N. Hincks}
\email{ihincks@uwaterloo.ca} 
\affiliation{Department of Applied Math, University of Waterloo, Waterloo, ON, Canada}
\affiliation{Institute for Quantum Computing, Waterloo, ON, Canada}
\author{D. G. Cory}
\affiliation{Institute for Quantum Computing, Waterloo, ON, Canada}
\affiliation{Perimeter Institute for Theoretical Physics, Waterloo, ON, Canada}
\affiliation{Department of Chemistry, University of Waterloo, Waterloo, ON, Canada}
\author{C. Ramanathan}
\affiliation{Department of Physics and Astronomy, Dartmouth College, Hanover, USA}

\begin{abstract}
Given a fixed initial state, a desired Hamiltonian, and an amount of time, we provide a complete
characterization of the set of Hamiltonians which perform the same action as the desired Hamiltonian
on the state of interest. An example is provided where the desired Hamiltonian lies outside of the
control Lie algebra, but implementable Hamiltonians exist within the characterization.
\end{abstract}

\maketitle

In addition to not always being feasible, having complete control over a quantum system is not always necessary.
Indeed, examples of this situation are numerous, and include quantum simulators designed to simulate only certain tasks~\cite{somaroo_quantum_1999,buluta_quantum_2009}, sensors which exploit coherent dynamics to measure in the nanoscale regime~\cite{cappellaro_entanglement_2005,taylor_high-sensitivity_2008}, and most protocols in quantum communication~\cite{bennett_teleporting_1993,mattle_dense_1996,bennett_quantum_1984}.
If one has a quantum circuit and wishes to implement it, a quantum system over which one has complete control would of course work, but a quantum device capable of implementing only one's circuit would nonetheless suffice, and perhaps in a more cost effective manor.
However, many of the tools of quantum control theory are targeted towards the cases of complete controllability, pure state controllability, and eigenstate controllability, which all emphasize a need to be able to control a given quantum system arbitrarily~\cite{albertini_lie_2002,dalessandro_introduction_2007,turinici_wavefunction_2003,zhang_control_2005}.
This paper develops a new tool to add to the toolbox of state to state map finding in the case of only partial quantum controllability.

The following is our predicament: we wish to evolve some state under a desired Hamiltonian which lies outside of our control algebra of implementable Hamiltonians for a specified time.
The goal is to give a full characterization of the Hamiltonians which perform the same action as our desired Hamiltonian on our state of interest.
Moreover, we will provide a physically motivated example where such an alternate Hamiltonian exists, and lies within our system's control.

Let $\Hilbert = \mathbb{C}_{\dHilbert}$ denote our system's Hilbert space of dimension $\dHilbert$.
Then the space of linear operators acting on $\Hilbert$ can be taken to be the $\dHilbert\times\dHilbert$ complex matrices $\linearOp{\Hilbert}=\genLin{\dHilbert}$ \footnote{We will use the standard convention of Lie theory, denoting Lie groups with upper-case symbols (such as $\GenLin{\dHilbert}$, the general linear group), and their Lie algebras with the same letters in lower-case gothic print (such as $\genLin{\dHilbert}$). \cite{faraut_analysis_2008}}.
Let the matrix $\RI$ denote our fixed initial state of interest, the matrix $\HDesire$ our desired Hamiltonian, the matrix $\HInternal$ the internal Hamiltonian of our system, and $\tnot$ the amount of time we are interested in evolving for.
Thus, our desired final state is given by $\RF:=\UDesire{\RI}\UDesire^\dagger$ where $\UDesire = e^{-i\tnot\HDesire}$.
Our control algebra, which will be denoted by $\LieAC$, is a real Lie subalgebra of $\linearOp{\Hilbert}$, and moreover, a Lie subalgebra of the $\dHilbert\times\dHilbert$ skew-Hermitian matrices $\skewHerm{\dHilbert}$.
Since we are assuming that $\HDesire$ lies beyond our control, we know that $\LieAC$ must in fact be a proper subalgebra of $\skewHerm{\dHilbert}$.

\begin{figure}
	\centering
	\includegraphics[width=\columnwidth]{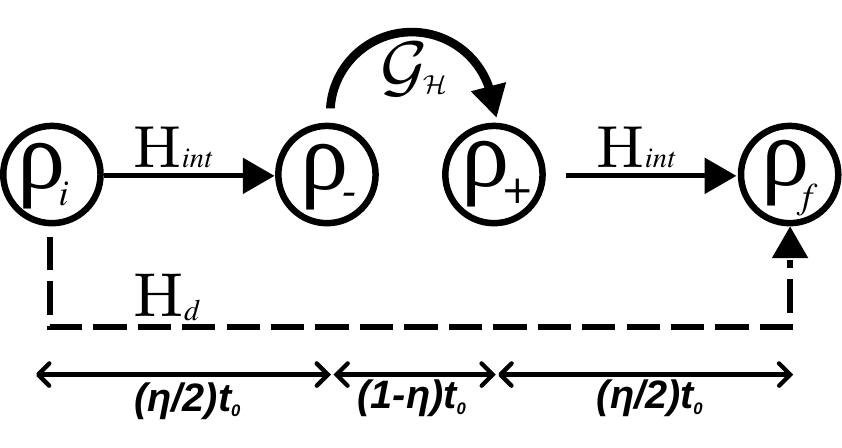}
	\caption{The initial state evolves under $\HInternal$ for a total time $\eta\tnot$, with an evolution under some member of $\GH$ for a time $(1-\eta)\tnot$ in between. This is arranged such that the state has effectively evolved under $\HDesire$ for a time $\tnot$.}
	\label{fig:jump}
\end{figure}

Fixing a fraction $0\leq\eta<1$, our scheme (illustrated in Figure~\ref{fig:jump}) will be to evolve under the internal Hamiltonian for a time $\frac{\eta}{2}\tnot$, perform control operations for a time $(1-\eta)\tnot$, and then resume evolution under the internal Hamiltonian for the remaining amount of time $\frac{\eta}{2}\tnot$.
The purpose the the parameter $\eta$ is to facilitate the case where the desired Hamiltonian is similar to the internal Hamiltonian, so that only a brief control sequence surrounded by natural evolution is necessary.
An example of this might be removing next-nearest neighbour couplings in a chain of spins, desiring only the effects of nearest neighbour interaction.
It can, of course, be set to zero if it is not needed.

We will denote the state of our system prior to the control sequence by $\RM$, and the state immediately following the control sequence by $\RP$, so that
\begin{align}
	\RM &= \conj{\RI}{\UInternal} \\
	\RP &= \jnoc{\RF}{\UInternal}
\end{align}
where $\UInternal=e^{-i\frac{\eta}{2}\tnot\HInternal}$.
Those Hamiltonians which evolve $\RM$ into $\RP$ in time $(1-\eta)\tnot$ constitute a set which we will call $\GH$.
Our method will be to first characterize the set $\GU$ of unitary matrices which take $\RM$ to $\RP$.
When this is done, $\GH$ will be found by taking all logarithms of all elements of $\GH$.
We may write
\begin{align}
	\GU &= \{U \in \Unitary{\dHilbert} ~|~ \RP = U \RM U^\dagger\} \\
	\GH &= \{H \in i\skewHerm{\dHilbert} ~|~ e^{-i\tnot(1-\eta){H}} \in \GU \}
\end{align}
where $\Unitary{\dHilbert}$ and $i\skewHerm{\dHilbert}$ are the $\dHilbert\times\dHilbert$ unitary matrices and Hermitian matrices respectively.

The states $\RM$ and $\RP$ being unitarily equivalent, we may diagonalize them to the same matrix as follows.
Let $\UM$ be a unitary matrix which diagonalizes $\RP$ to the diagonal matrix $D$, so that $\RM = {\UM}{D}{\UM^{\dagger}}$.
Then $\UP := \UInternal^\dagger\UDesire\UInternal^\dagger\UM$ is a unitary matrix which diagonalizes $\RP$ to the same diagonal matrix $D$.

In general, we can think of a diagonal matrix as a block diagonal matrix where each block is a scalar multiple of an identity matrix and no two adjacent blocks share their diagonal value in common.
If $T$ is a diagonal matrix, let lower-case $t_j$ be the scalar value corresponding the $j^\text{th}$ block, $p_{j}^{T}$ be the size of the $j^\text{th}$ block, and $m_T$ be the number of blocks.
For example, this gives us $D = \diagb{d_{j}I_{p_{j}^{D}}}{j=1}{m_D}$. 

Without loss of generality, by swapping the columns of $\UM$ and $\UP$, we may suppose that $D$ has been chosen so that $d_i \neq d_j$ whenever $i \neq j$.
Now $\RM$ and $\RP$ share the same set of eigenvalues, namely $\{d_j\}_{j=1}^{M_D}$, and with the same degeneracies $\{p_{j}^{D}\}_{j=1}^{m_D}$.

We begin our characterization of $\GU$ and $\GH$ with an easily verified matrix property. Suppose $X \in \GenLin{n}$ and that $T$ is a diagonal matrix such that $t_i \neq t_j$ whenever $i \neq j$.
Then $T = XTX^{-1}$ if and only if $X \in \GenLinBlock{p_{1}^{T}}{p_{m_T}^{T}}$~\footnote{$\GenLinBlock{d_1}{d_n}$ will denote the invertible block diagonal $\dHilbert\times\dHilbert$ matrices with block sizes $d_1,d_2,\dots,d_n$. Other block diagonal structures will share similar notation.}.

As a result, we conclude that if $U \in \mathcal{U}(\dHilbert)$ is a unitary matrix, then $D = U{D}U^\dagger$ if and only if $U \in \UnitaryBlock{p_{1}^{D}}{p_{m_D}^{D}} \subseteq \mathcal{U}(\dHilbert)$, from which $(*)$ in the following string of equivalences follow:
\begin{align*}
	& W \in \GU  \iff \RP = W \RM W^\dagger \\
			  & \iff \UP D \UP^\dagger = W (\UM D \UM^\dagger) W^\dagger \\
			  & \iff D = \UP^\dagger W \UM D \UM^\dagger W^\dagger \UP \\
			  & \stackrel{(*)}{\iff} U:= \UP^\dagger W \UM \in \UnitaryBlock{p_{1}^{D}}{p_{m_D}^{D}} \\
			  & \iff W = \UP U \UM^\dagger \text{, with } U \in \UnitaryBlock{p_{1}^{D}}{p_{m_D}^{D}}
\end{align*}
And so our characterization of $\GU$ is just a left and right translation of the block diagonal group $\UnitaryBlock{p_{1}^{D}}{p_{m_D}^{D}}$ by the unitaries $\UP$ and $\UM^\dagger$ respectively:
\begin{equation}
	\GU = \UP\UnitaryBlock{p_{1}^{D}}{p_{m_D}^{D}}\UM^\dagger
\end{equation}

The next step is to characterize the set $\GH$ of Hamiltonians which carry us from $\RM$ to $\RP$.
The difficulty in this step arises from the non-injectivity of the matrix exponential function.
For each $U \in \GU$ there are at least countably many valid matrix logarithms, and uncountably many if $U$ has any degenerate eigenvalues.

For any $U \in \GU$ let $\DU$ be the set of ordered pairs describing the diagonalizations of $U$, i.e., define 
\begin{equation}
	\DU = \left\{(V,T)~ 
	\begin{array}{|l}
		U = VTV^{\dagger} \text{, where } \\
		V \in \mathcal{U}(d_{\Hilbert}) \text{ and } \\
		T \text{ is diagonal}
	\end{array}
	\right\}
\end{equation}
Since all of the $T$'s in $\DU$ are equal up to permutation, we can use the matrix property stated above in a similar way to derive the following more useful description of $\DU$
\begin{equation}
\DU =
\left\{
\left(V'WP,~P^{\dagger}T'P\right)
\begin{array}{|l}
P \text{ a permutation matrix} \\
W \in \UnitaryBlock{p_{1}^{T'}}{p_{}^{T'}}
\end{array}
\right\}
\end{equation}
where $(V',T') \in \DU$ was chosen so that ${t'}_i \neq {t'}_j$ whenever $i \neq j$.

With all of this notation aside, we can start to talk of logarithms.
Given some $U \in \GU$, all of the logarithms of $U$, that is, all of the matrices (skew-Hermitian or otherwise) which exponentiate to $U$, are of the form
\begin{equation}
\label{eq:logform}
A = VX_{T}\diagb{i\phi_{j}+2\pi{i}{k_j}}{j=1}{\dHilbert}X_{T}^{-1}V^{\dagger}
\end{equation}
where $(V,T) \in \DU$, $T = \diagb{e^{i\phi_j}}{j=1}{\dHilbert}$, and $X_T$ is an invertible matrix which commutes with $T$, i.e. $X_T$ lies in the general linear centralizer of T, $\centralizer{T}$.
It is easy to check that a matrix of this form exponentiates to $U$:
\begin{align}
	e^{A} & = e^{VX_{T}\diag{i\phi_{j}+2\pi{i}{k_j}}X_{T}^{-1}V^{\dagger}} \notag \\
	& = V\underbrace{X_{T}\diag{e^{i\phi_{j}}}}_\text{commute}X_{T}^{-1}V^{\dagger} 
	= U
\end{align}

For a proof that the form in Equation~\ref{eq:logform} constitutes \emph{all} such logarithms, the reader is redirected to Reference~\cite{gantmacher_theory_1998}.
We finally arrive at the following characterization of $\GH$:

\begin{widetext}
\begin{equation}
\label{eq:characterization}
\GH =
\left\{
\frac{1}{(1-\eta)\tnot}
VX_{T}\diagb{-\phi_{j}+2\pi{k_j}}{j=1}{\dHilbert}X_{T}^{-1}V^{\dagger}
\begin{array}{|l}
U \in \GU \text{, } (V,T) \in \DU \text{ with } \\
T=\diagb{e^{i\phi_{j}}}{j=1}{\dHilbert} \text{, and} \\
X_T \in \centralizer{T} \text{, } k_j \in \mathbb{Z}
\end{array}
\right\}
\cap i\skewHerm{\dHilbert}
\end{equation} 
\end{widetext}

Note that we have multiplied the form in Equation~\ref{eq:logform} by $(-i(1-\eta)\tnot)^{-1}$.
This is simply so that if $H\in\GH$, then $e^{-i(1-\eta)\tnot H}=U$ rather than $e^H = U$.

To see that the intersection with $i\skewHerm{\dHilbert}$ is necessary, in other words, to see that a logarithm of a unitary is not necessarily skew-Hermitian, observe that our characterization of logarithms tells us that
$A=\bigl(\begin{smallmatrix}1&1\\ 0&1\end{smallmatrix} \bigr)\bigl(\begin{smallmatrix}0&0\\ 0&2\pi{i}\end{smallmatrix} \bigr)\bigl(\begin{smallmatrix}1&1\\ 0&1\end{smallmatrix} \bigr)^{-1}$ exponentiates to the identity matrix.
But $A = \bigl(\begin{smallmatrix}2\pi{i}&-2\pi{i}\\ 0&0\end{smallmatrix} \bigr)$, which is clearly not skew-Hermitian.

If two matrices are unitarily equivalent, then the first is Hermitian if and only if the second is Hermitian.
Thus a sufficient condition for the matrix $A = VX_{T}\diag{-\phi_{j}+2\pi{k_j}}X_{T}^{-1}V^{\dagger}$ to be Hermitian is for the matrix $X_T$ to be unitary, so that the Hermitian matrix $\diag{-\phi_{j}+2\pi{k_j}}$ would be unitarily equivalent to $A$ via the unitary $VX_T$.
This condition is not in general necessary.
The necessary and sufficient condition is that $X_{T}^{\dagger}X_{T}$ commutes with $\diag{-\phi_{j}+2\pi{k_j}}$, which can easily be derived, once more using the stated matrix property.
Hence, the intersection in Equation~\ref{eq:characterization} can be removed if the statement $[X_T^\dagger X_T, \diag{-\phi_{j}+2\pi{k_j}}]=0$ is appended to the conditions.

We may count that there are four freedoms at our disposal when picking a member of $\GH$.
The first comes from our characterization of $\GU$ and is the choice of any block-diagonal unitary $U$ in the group $\UnitaryBlock{p_{1}^{D}}{p_{m_D}^{D}}$, whose dimension as a manifold is determined by the amount of degeneracy in the eigenvalues of the state $\RM$ (or equivalently, $\RP$).
This dimension can range from $(\dHilbert-1)^2$, when $\RM$ is a pure state, to $\dHilbert$, when no two of $\RM$'s eigenvalues are equal.

Once $U$ has been chosen, the second freedom lies in the choice of diagonalization, $(V,T)$, of $U$, which is characterized by the set $\DU$.
As we have seen, choosing $(V,T)$ comes down to choosing a unitary in $\UnitaryBlock{p_{1}^{T'}}{p^{T'}}$, as well as any permutation matrix $P$.
Thus the size of $\DU$ is determined by the amount of degeneracy in the eigenvalues of $U$.

The third freedom comes in the form of $\dHilbert$ choices of integers for $k_1$ through $k_{\dHilbert}$.

The final freedom is in the choice of member from the centralizer group $\centralizer{T}$.
The size of this freedom is dependant on the number of unique values contained in $T$, i.e. the degeneracy of $U$ once again.
We must also remember that this freedom is restricted if we wish for our Hamiltonians to be Hermitian.

We now turn toward an illuminating example.
Recall that $\LieAC$ is defined to be the Lie algebra generated by our system controls.
In the case where $\GH$ and $i\LieAC$ have empty intersection, we simply do not have enough control to do as we wish~\footnote{Because the algebra $\LieAC$ contains skew-Hermitian matrices, we will multiply it by the complex unit $i$ when we wish to compare it with collections of Hermitian matrices}.
On the other hand, if $\GH \subseteq i\LieAC$, then we have too much control for our scheme to be useful; we can evolve under $H_d$ directly.
And so for us, the only interesting case is when $\GH$ and $i\LieAC$ intersect, but neither set is contained within the other.
In this case, perhaps, $H_d$ lies outside of $i\LieAC$, but we can effectively simulate the piece we are interested in for a specified time $t_0$ with some $H \in i\LieAC\cap\GH$.

It is not immediately clear that this third and most interesting case should ever happen.
Perhaps each $H$ in $\GH$ is implementable if and only if one $H$ in $\GH$ is implementable.
The following provides a counterexample to this idea.

Suppose we are interested in passing information from one end of a four qubit spin chain to the other, and that this spin chain's internal Hamiltonian is given by a dipole-dipole interaction
\begin{equation}
	\HInternal = \sum_{1 \leq k < l \leq 4} J_{kl}(\PX{k}\PX{l} + \PY{k}\PY{l} - 2\PZ{k}\PZ{l})
\end{equation}
where $J_{kl}\propto |k-l|^{-3}$ falls off as distance cubed. 

Suppose also that we have some control over the system, the global $X$ and $Y$ rotations whose Hamiltonians are given by
\begin{equation}
	H_{x} = \sum_{k=1}^{4} \PX{k} \text{ and } H_{y} = \sum_{k=1}^{4} \PY{k}
\end{equation}

The propagation of quantum information along spin chains in the case of either no or limited control has been well studied~\cite{bose_quantum_2008}.
The simplest method, which naturally yields poor fidelities, is to allow the free evolution of an $XY$ nearest-neighbour Hamiltonian with uniform couplings and wait for a specific amount of time~\cite{bose_quantum_2003}.
If one has the ability to engineer non-uniform nearest-neighbour couplings, then the same idea can produce unit fidelity transfers~\cite{christandl_perfect_2004,wang_all_2011}.
Again in the case of no control on the spin chain, if one has a dual rail spin chain, or equivalently, a single rail of qutrits, unit transfer fidelity of a qubit can be achieved with a wide class of chain Hamiltonians~\cite{burgarth_conclusive_2005}.
If one adds the ability to perform certain controls to a nearest-neighbour chain, then pulse sequences can be found on uniformly coupled models with unit fidelity transfers~\cite{fitzsimons_globally_2006}, and with engineered couplings the state of the chain doesn't even need to be initialized~\cite{di_franco_perfect_2008}.
The example presented here excels in that the chain's interaction is not restricted to nearest-neighbour couplings, and that the controls are only required to be global rotations, but lags in its numerical tractability for long chains.

We take the following $XY$ Hamiltonian as our desired Hamiltonian
\begin{equation}
	H_{XY} = \sum_{k=1}^{4-1} C_{k}(\PX{k}\PX{k+1} + \PY{k}\PY{k+1})
\end{equation}
because if the coefficients are chosen as $C_{k} = \frac{\lambda}{2}\sqrt{k(4-k)}$, then information can be passed from the first spin on the chain to the last spin on the chain with unit fidelity simply by waiting for a time $\pi/\lambda$\cite{christandl_perfect_2004}.
In other words, for any $\alpha$ and $\beta$ we have the following property:
\begin{align}
	\label{eq:transfer}
	e^{-i\frac{\pi}{\lambda}H_{XY}} &(\alpha\ket{\pa} + \beta\ket{\pb}) \notag \\ &= \alpha\ket{\pc} + \beta\ket{\pb}
\end{align}

By taking nested commutators of our three control Hamiltonians ($H_x$, $H_y$, and $H_{DD}$) until all further nested commutators lie in the span of the previous ones, we find that our Lie algebra has a dimension of $96$, with the dimension of the full control space being $4^{4} = 256$~\cite{dalessandro_introduction_2007}.
The norm distance between $H_{XY}$ and its projection onto the space $i\LieAC$ can be computed to be non-zero, so that $H_{XY}$ is not in our control algebra.
This means that if a solution is found to our problem, it will be a non-trivial one.

Let our initial state be the pure state $\RI = \ket{\pa}\bra{\pa}$, and our fraction of internal evolution be $\eta = 0.95$.
We choose the diagonalization unitary $\UM$ so that $D = \diag{1,0,\cdots,0}$, hence $\GU = \UP\mathcal{U}(1,15)\UM^\dagger$.

We wish for our solution to retain the property in Equation~\ref{eq:transfer}, but in choosing a unitary $U$ from $\GU$, we can only be guaranteed that this property holds for one state, namely, $\UInternal{U}\UInternal\ket{\pa} = \ket{\pc}$.
To overcome this limitation, we use the following trick. We choose a subspace of $\LieAC$, denoted $\LieAC(\down)$, such that each matrix in this subspace has $\ket{\pb}$ as an eigenvector.
We can expect such a subspace to exist since $\ket{\pb}$ is an eigenvector of both $\HInternal$ and $[H_x,H_y]$.
In our calculations, such a subspace was found having a dimension of $9$, where no attempt was made to find the largest such subspace.
Now if we were to successfully perform the optimization
\begin{equation}
	\label{eq:costfunction}
	\HCalc = \argmin_{H \in \GH}
	\Vert H - \project{H}{i\LieAC(\down)} \Vert
\end{equation}
where $i\mathcal{P}_{\LieAC(\down)}$ is the projection operator onto $i\LieAC(\down)$, we would be ensured that
\begin{equation}
	\UInternal e^{-i(1-\eta)\tnot{\HCalc}}\UInternal\ket{\pb} = \ket{\pb}
\end{equation}
and hence by linearity, and since $\HCalc\in\GH$, the desired property
\begin{align}
	\UInternal &e^{-i(1-\eta)\tnot{\HCalc}}\UInternal (\alpha\ket{\pa} + \beta\ket{\pb}) \notag \\ &= \alpha\ket{\pc} + \beta\ket{\pb}
\end{align}
would be recovered for any $\alpha$ and $\beta$.

Actually performing this minimization is non-trivial since the spaces we must optimize over are non-linear, and moreover, a full optimization would require a double optimization, first over $\GU$ and then over $\DU$, $\mathbb{Z}^{\dHilbert}$ and $\centralizer{T}$.
However, in our example a partial optimization turns out to suffice, where we optimize only over $\GU$, leaving our other freedoms fixed at simple values ($k_j=0$ for all $j$, $X_T=I$, and letting Matlab's diagonalization function choose the member of $\DU$).

\begin{figure}
	\centering
	\includegraphics[width=\columnwidth]{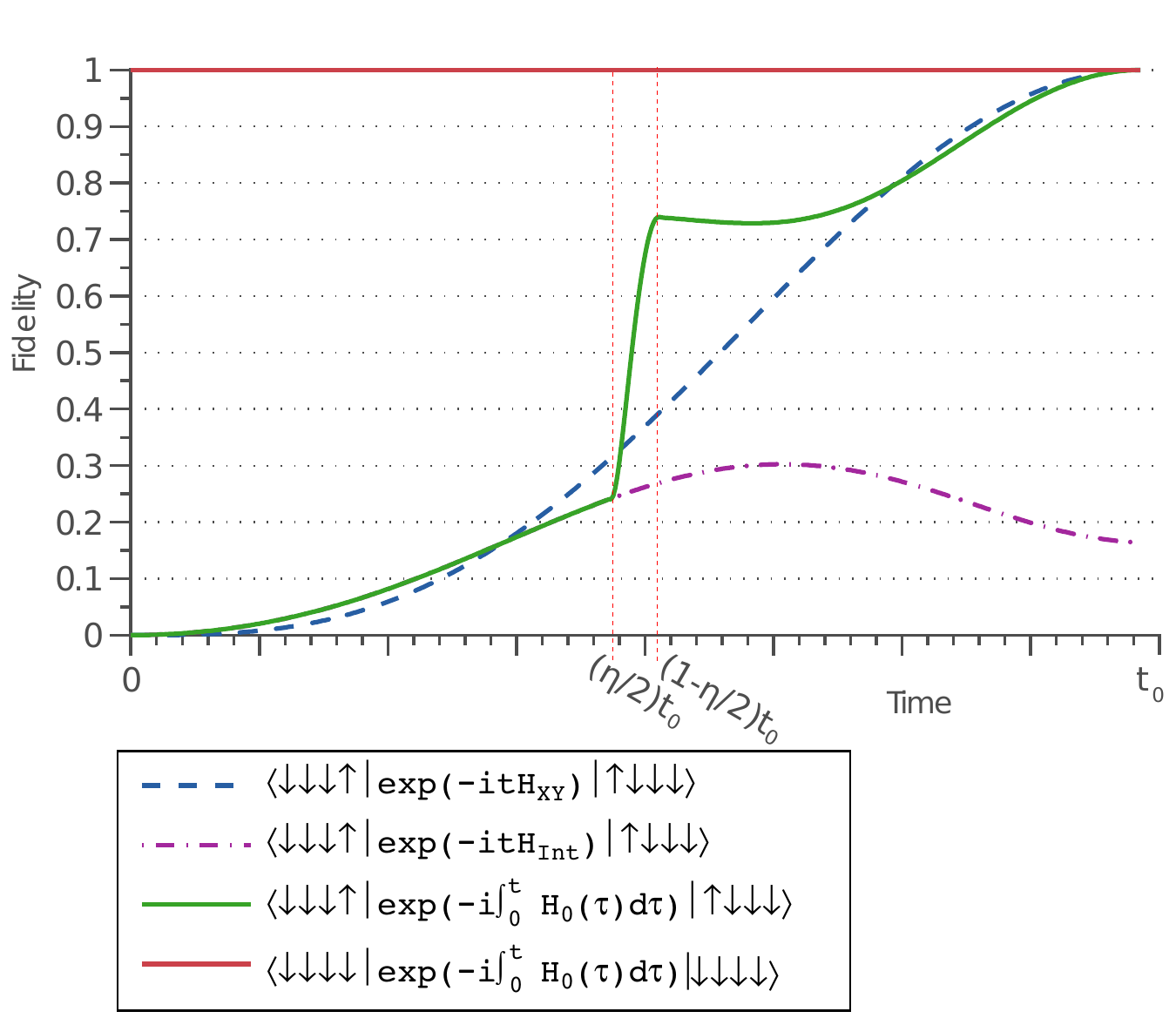}
	\caption{The fidelity of information transfer from the first to the last qubit on the spin chain for the states of interest, $\ket{\pa}$ and $\ket{\pb}$, evolving under the numerically discovered Hamiltonian $H_0$. Also included for comparison are the fidelities of transfer under the $H_{XY}$ and $H_{int}$ Hamiltonians. Recall that $\ket{\pb}$ is an eigenvector of all Hamiltonians involved.}
	\label{fig:fidelities}
\end{figure}

This scheme yields a Hamiltonian $\HCalc$ which lies in both $\GH$ and $i\LieAC(\down)$, and obtains the global minimum of $0$ for the cost function written down Equation~\ref{eq:costfunction}.
It should be noted that both of the maximizations
\begin{align}
	\HCalc^{\prime}  = \argmax_{H \in \LieAC(\down)} | &
	\bra{\pc}e^{-i(1-\eta)\tnot H}\ket{\pa}| \\
	\HCalc^{\prime\prime}  = \argmax_{H \in \LieAC} ( | &
	\bra{\pc}e^{-i(1-\eta)\tnot H}\ket{\pa}| \\ 
	+ |&\bra{\pb}e^{-i(1-\eta)\tnot H}\ket{\pb}|)
\end{align}
would be much simpler, and produce effectively the same result in this example. The difference, however, is that these two methods demand that our operation be implementable while seeking a maximum in state transfer fidelity, whereas the minimization in Equation~\ref{eq:costfunction} demands
that the fidelity be perfect while attempting to minimize the distance to implementability.

Figure \ref{fig:fidelities} shows the fidelity of state transfer as a function of time, where $H_0$ is the piecewise constant Hamiltonian which begins and ends as $\HInternal$, and is equal to $\HCalc$ for a duration $(1-\eta)\tnot$ in between.
Similar results have been computed for both $5$ and $6$ qubits.

Having calculated the Hamiltonian $\HCalc$, the methods of optimal control theory can be used to generate control sequences which implement it~\cite{khaneja_optimal_2005}. If desired, the control sequence can subsequently be concatenated with its time and phase reversal to generate a universal rotation gate~\cite{luy_construction_2005}.

In conclusion, we have provided a complete characterization of the set of Hamiltionians which simulate the evolution of a given density matrix under a desired, but not implementable, Hamiltonian.
Moreover, the above example demonstrates that in a non-trivial physically motivated scenario, it is possible for this set to have nonempty intersection with the control algebra, thereby providing a realizable means of doing what can not be done directly.

\emph{Acknowledgements -} This work was supported by the Canadian Excellence Research Chairs (CERC) program.

\bibliography{mybib}

\end{document}